\documentclass[twocolumn]{aastex63}

\newcommand\aastex{AAS\TeX}

\newcommand{\sigpr}{\sigma'_{\star,\rm{int}}}
\newcommand{\sigprsq}{\sigma'^2_{\star,\rm{int}}}
\usepackage{enumerate} 

\usepackage{import}
\usepackage{amsmath}
\usepackage{amssymb}
\shorttitle{\aastex\ Virial Mass Estimates}
\shortauthors{van der Wel et al.}

\begin{document}

\title{The Mass Scale of High-Redshift Galaxies: Virial Mass Estimates Calibrated with Stellar Dynamical Models from LEGA-C}

\correspondingauthor{Arjen van der Wel}
\email{arjen.vanderwel@ugent.be}

\author{Arjen van der Wel}
\affil{Sterrenkundig Observatorium, Universiteit Gent, Krijgslaan 281 S9, 9000 Gent, Belgium}

\author[0000-0002-0786-7307]{Josha van Houdt}
\affil{Max-Planck Institut f\"{u}r Astronomie K\"{o}nigstuhl, D-69117, Heidelberg, Germany}

\author{Rachel Bezanson}
\affil{University of Pittsburgh, Department of Physics and Astronomy, 100 Allen Hall, 3941 O'Hara St, Pittsburgh PA 15260, USA}

\author{Marijn Franx}
\affil{Leiden Observatory, Leiden University, P.O.Box 9513, NL-2300 AA Leiden, The Netherlands}

\author{Francesco D'Eugenio}
\affil{Sterrenkundig Observatorium, Universiteit Gent, Krijgslaan 281 S9, 9000 Gent, Belgium}

\author{Caroline Straatman}
\affil{Sterrenkundig Observatorium, Universiteit Gent, Krijgslaan 281 S9, 9000 Gent, Belgium}

\author{Eric F.~Bell}
\affil{Department of Astronomy, University of Michigan, 1085 South University Ave., Ann Arbor, MI 48109, USA}

\author{Adam Muzzin}
\affil{Department of Physics and Astronomy, York University, 4700 Keele St., Toronto, Ontario, M3J 1P3, Canada}

\author{David Sobral}
\affil{Department of Physics, Lancaster University, Lancaster LA1 4YB, UK}

\author{Michael V.~Maseda}
\affil{Leiden Observatory, Leiden University, P.O.Box 9513, NL-2300 AA Leiden, The Netherlands}

\author{Anna de Graaff}
\affil{Leiden Observatory, Leiden University, P.O.Box 9513, NL-2300 AA Leiden, The Netherlands}

\author{Bradford P.~Holden}
\affil{UCO/Lick Observatory, University of California, Santa Cruz, CA 95064, USA}

\begin{abstract}
Dynamical models for $673$ galaxies at $z=0.6-1.0$ with spatially resolved (long-slit) stellar kinematic data from LEGA-C are used to calibrate virial mass estimates defined as $M_{\rm{vir}}=K \sigprsq R$, with $K$ a scaling factor, $\sigpr$ the spatially-integrated stellar velocity second moment from the LEGA-C survey and $R$ the effective radius measured from a S\'ersic profile fit to HST imaging. The sample is representative for $M_{\star}>3\times10^{10}~M_{\odot}$ and includes all types of galaxies, irrespective of morphology and color.  We demonstrate that using $R=R_{\rm{sma}}$~(the semi-major axis length of the ellipse that encloses 50\% of the light) in combination with an inclination correction on  $\sigpr$~produces an unbiased $M_{\rm{vir}}$.
We confirm the importance of projection effects on $\sigpr$ by showing the existence of a similar residual trend between virial mass estimates and inclination for the nearby early-type galaxies in the ATLAS$^{\rm{3D}}$~survey.
Also, as previously shown, when using a S\'ersic profile-based $R$ estimate, then a S\'{e}rsic index-dependent correction to account for non-homology in the radial profiles is required. 
With respect to analogous dynamical models for low-redshift galaxies from the ATLAS$^{\rm{3D}}$~survey we find a systematic offset of 0.1 dex in the calibrated virial constant for LEGA-C, which may be due to physical differences between the galaxy samples or an unknown systematic error. Either way, with our work we establish a common mass scale for galaxies across 8 Gyr of cosmic time with a systematic uncertainty of at most 0.1 dex.
\end{abstract}


\section{Introduction}
\label{section:intro}
The total mass of a galaxy is perhaps the single most important of its properties. At all cosmic times, it is related to a host of other properties, such as the stellar mass \citep[e.g.,][]{Taylor2010}, star formation rate (\citealt{Brinchmann2004, Noeske2007}), size \citep[e.g.,][]{Shen2003, VanderWel2014a}, average stellar age (e.g. \citealt{Gallazzi2005, Gallazzi2014, Wu2018}) and rotational properties (e.g., \citealt{Emsellem2007}). Furthermore, a galaxy can be plausibly linked to its progenitors through its mass, which changes through time due to passive growth or mergers \citep[e.g.,][]{Bezanson2009, Naab2009}. Deriving accurate, unbiased masses is therefore an obvious priority in any galaxy survey.

For spatially integrated kinematic measurements we define the virial mass using the scalar virial theorem:

\begin{equation}
    M_{\rm{vir}} = K\frac{\sigma^{2}R}{G}
    \label{eq:virial_mass}
\end{equation}

\noindent
with $G$ the gravitational constant, $R$ the radius of the galaxy, $\sigma$ the velocity second moment, strictly speaking, only equal to the velocity dispersion for a non-rotating galaxy, and $K$ a scaling factor. All of these quantities, with the exception of $G$,  are at this point only generically defined and their exact definition is, in essence, the topic of this paper.

The parameters used in this $M_{\rm{vir}}$ estimate only crudely approximate the kinematic and geometric structure of galaxies and are therefore susceptible to both random and systematic uncertainties. Two approaches have been used to address this issue: comparison with dynamical models, and comparison with stellar mass estimates from photometry.  First, dynamical models provide an accurate absolute mass scale with which $M_{\rm{vir}}$ estimates can be compared. Such an empirical calibration was done for the SAURON survey in \citet{Cappellari2006} and revisited by the ATLAS$^{\rm 3D}$ collaboration in \citet{Cappellari2013+jam} (hereafter C06 and C13, respectively).
This volume-limited sample of morphologically selected early-type galaxies produces a large dynamic range in mass, but the number of high-mass galaxies is relatively small. Larger surveys across all galaxy types (\citealt{CALIFA2012,MANGA2015,SAMI_DR2}) have not revisited this calibration, even though this would be relatively straightforward with the dynamical models for 2000 galaxies in MANGA at $z\lesssim 0.1$ published by  \citet{Li+2018}.

Second, a comparison between $M_{\rm{vir}}$ and stellar mass estimates $M_*$ provides an idea of the precision of both parameters (\citealt{Taylor2010}). The average value and scatter in $M_*/M_{\rm{vir}}$ are informative, but both quantities suffer from systematics, again leaving the absolute mass scale uncertain. This method is popular at higher redshifts (\citealt{VdWel2006,VdSande2013,Belli2014}) where (until now) dynamical models have been difficult to construct, resulting in small and biased samples of old, massive galaxies (\citealt{DokkumMarel2007,WelMarel2008, Shetty2015,Guerou2017,Newman2018}). 

In this paper we make progress in addressing these issues by using nearly 800 galaxies in the $z=0.6-1$ redshift range from the LEGA-C survey (\citealt{LEGAC2016, vanderwel21}) with dynamical models from \citet[][hereafter, vH21]{houdt21}. This sample includes all types of galaxies as the selection was blind to structure and color.  Using mass estimates from the dynamical models we will establish the normalization $K$, show that the use of a circularized radius should be avoided, and introduce a necessary but simple inclination correction for the integrated velocity second moment. We assume a $\Lambda$CDM cosmology with $H_{0}=70$ kms$^{-1}$, $\Omega_{\Lambda}=0.7$ and $\Omega_{M}=0.3$.

\section{Data \& Methods} 
\label{section:data}

This work is based on the Large Early Galaxy Astrophysics Census survey (LEGA-C; the survey description and further details can be found in \citealt{LEGAC2016,LEGAC2018,vanderwel21}). This survey provides exceptionally deep, spatially resolved spectra for approximately $4000$ magnitude-limited galaxies from the UltraVISTA photometric parent catalog (\citealt{UltraVista2013}), targeted at redshifts between $0.6$ and $1.0$. Spectra have been obtained with the VIMOS instrument on the Very Large Telescope. With $\sim20$ hours of integration per object, $R\sim3500$ spectra are produced with a wavelength coverage between $\sim6300$ and $\sim8800$\AA.

In this paper, we use the subset of galaxies drawn from the third data release \citep[DR3;][]{vanderwel21} suitable for kinematic modeling described in full by \citet{houdt21}. Galaxies are selected to have high signal-to-noise ratio ($>10$\AA$^{-1}$), measured $\sigpr$ (see below for definition), have major axes aligned within $45$ degrees of the direction of the slit, and have imaging data show a regular morphology that is well described by a S\'{e}rsic profile ({\tt FLAG\_MORPH}~$=0$ in DR3). 

Stellar- and gas kinematics are derived from the spectra using pPXF\footnote{v6.0.0, via \url{http://www-astro.physics.ox.ac.uk/~mxc/software/}} (\citealt{pPXF-2004,pPXF-2017}). In summary, a combination of single stellar population (SSP) templates and Gaussian emission lines are fit to the observed spectra. The theoretical spectra are broadened and shifted to find the spatially resolved rotation and dispersion, independently for the gas and stars. This is done for 2D and 1D spectra, where the former is used for the Jeans models (see below) and the latter is used to extract the integrated velocity second moment $\sigpr$ which are used to calculate the virial masses. For further details on the spectral modeling, see \citet{Bezanson2018,Bezanson2018b}. The notation $\sigpr$, introduced by \citet{Bezanson2018b}, is chosen to differentiate between the spatially integrated velocity broadening along the line of sight and the intrinsic velocity dispersion $\sigma_{\star}$.

Optical imaging is available for each galaxy with HST/ACS F814W observations (\citealt{COSMOS2007}). Structural parameters are extracted with single-component S\'{e}rsic fits using GALFIT\footnote{v3.0.5, available at \url{https://users.obs.carnegiescience.edu/peng/work/galfit/galfit.html}} (\citealt{GALFIT2010}) as described in \citet{VdWel2012} and DR3. All three main parameters from the S\'ersic fit -- S\'ersic index, effective radius, and projected axis ratio -- play a key role in this work and their use in Equation \ref{eq:virial_mass} is, essentially, the topic of this paper.

The dynamical masses to which the virial masses will be scaled were obtained from Jeans models as presented by \citet{houdt21}. Summarising, the galaxies are modelled as oblate axisymmetric spheroids as implemented in the Jeans Anisotropic Multi-Gaussian Expansion (JAM) code (\citealt{JAM2008})\footnote{v5.0.17 from https://pypi.org/project/jampy/}. The surface brightness is parameterised by the S\'{e}rsic profiles derived from the HST/ACS F814W imaging, decomposed into a series of Gaussians using the MGE\footnote{v5.0.12, from \url{https://pypi.org/project/mgefit/}} (\citealt{MGE2002}) code. The probability density of the inclination is assumed to be a function of the observed axis ratio, using observationally derived intrinsic shape distributions (\citealt{Chang2013,VdWel2014b,houdt21}). The slit geometry of the LEGA-C spectroscopy is included in the models: the Jeans equations are integrated through rectangular $1\arcsec \times 0.205\arcsec$ apertures, instrumental velocity gradients are subtracted, and the centering of the slit is marginalized over in the Bayesian fitting approach. The model predictions of $v_{\rm{rms}}$ are compared with the measured $\sqrt{v'^2_{\rm{\star}}+\sigma'^2_{\star}}$, where $v'_{\rm{\star}}$ and $\sigma'_{\star}$ are the measured line of sight velocity first and second moments for each spatial element, typically reached $1-1.5\arcsec$ or $1-3$ effective radii along the slit direction.

There are two components in the gravitational potential: the stellar component (for which we assume that mass follows light as seen in the HST image) and a dark matter component, parameterized by a NFW halo. We do not claim to constrain the dark matter mass directly, but the inclusion of a dark component is required by the data and allows greater flexibility in fitting a gradient in the mass-to-light ratio regardless of its origin (stellar $M/L$, stellar Initial Mass Function (IMF), gas, dark matter), and therefore produces more realistic uncertainties. 

In vH21 we already published mass estimates and proxies for dynamical structure (e.g., $V/\sigma$), but to facilitate comparisons with other datasets and models we present in Appendix A the fitted parameters and model components. In this paper we use the sample of 673 galaxies selected to have JAM mass estimates within $R_e$ with a precision better than 0.5 dex. This is 22\% of the sample of galaxies for which we can estimate virial masses based on the integrated $\sigma'_{\star}$, which is essentially the full LEGA-C primary, K-band selected sample described by \citet{vanderwel21}. As explained by vH21, many galaxies do not have a Jeans mass estimate because their major axis is misaligned with the LEGA-C slit by more than 45 degrees ($\sim 50$\%, given the random position angle distribution), while other galaxies do not have spectra with sufficient signal-to-noise. However, as we will discuss further below, we find no correlations between the virial-to-Jeans mass ratio and any other parameter, which implies that our derived virial mass estimates can be applied generally to galaxies in the mass and redshift range of LEGA-C, which samples galaxies with stellar masses $\gtrsim2\times10^{10}~M_{\odot}$ and $50\lesssim \sigma'_{\star}/ (\rm{km s}^{-1}) \lesssim 300$, at $0.6<z<1$.

\section{Calibration of the Virial Mass}
\label{section:results}
Conceptually, there are two critical aspects that need to be addressed when estimating virial masses based on integrated velocity dispersions (or, to be more precise, second moments) and effective radius measurements: how non-homology in radial structure affects the virial mass (discussed in Section \ref{sec:results_p2}) and how to take into account non-homology in the 3D geometry of galaxies and the resulting projection effects on the observables (discussed below in Section \ref{sec:results_p1}).  

\subsection{Non-Homology in Radial Structure}
\label{sec:results_p2}

\begin{figure}[!h]
\epsscale{1.15}
\plotone{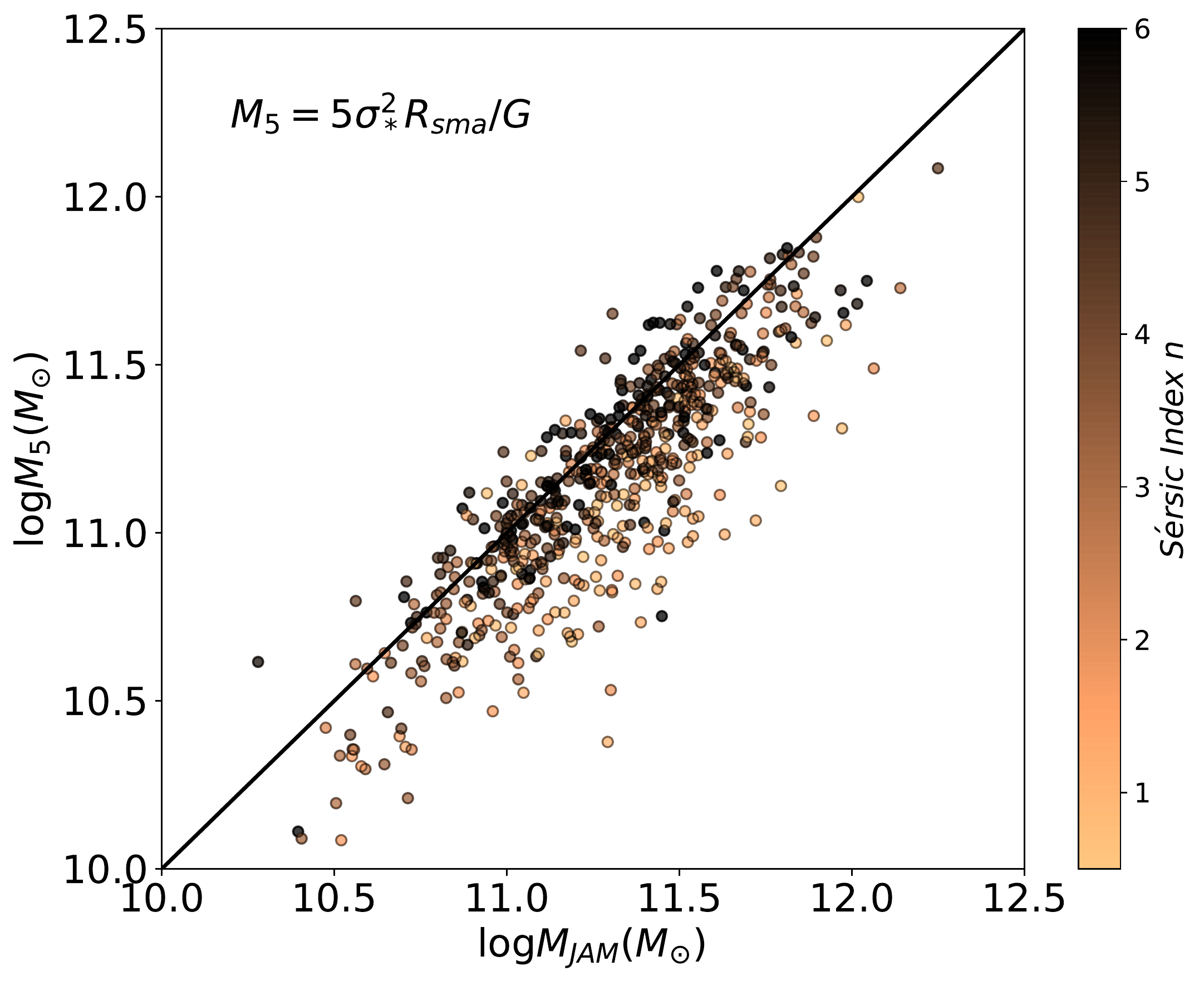}
\caption{The simplest virial mass estimator, $M_5$, versus $M_{\rm{JAM}}$, color-coded with S\'ersic index $n$. For low-$n$~galaxies $M_{\rm{JAM}}$ is systematically larger than $M_5$, implying that a non-homology correction is required to properly scale the virial mass estimator.
\label{fig:mass_mass}}
\end{figure}

\begin{figure}[!h]
\epsscale{1.15}
\plotone{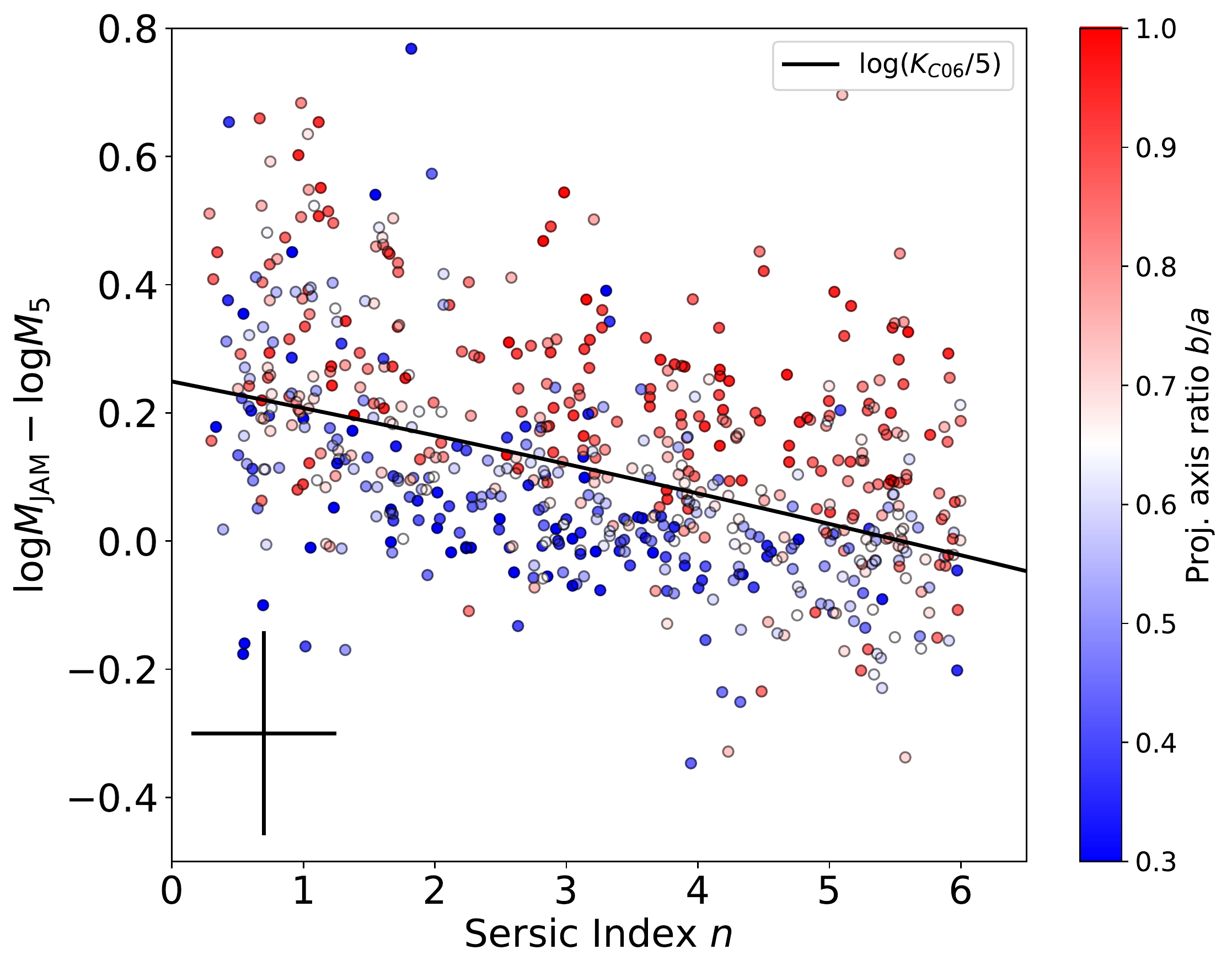}
\caption{Ratio of $M_{\rm{JAM}}$ and $M_5$~(see Fig.~\ref{fig:mass_mass}) versus S\'ersic index $n$, color-coded with projected axis ratio $b/a$. The black line represents the proportionality factor $K(n) = 8.87-0.831n+0.0241n^{2}$ from \citet{Cappellari2006} divided by a factor 5 (from $M_5$), which describes the $n$ dependence well. But a strong residual correlation with projected axis ratio is readily apparent, implying that a viewing angle correction is required to properly scale the virial mass estimator. \label{fig:n_k}}
\end{figure}

\begin{figure*}[t]
\epsscale{1.2}
\plotone{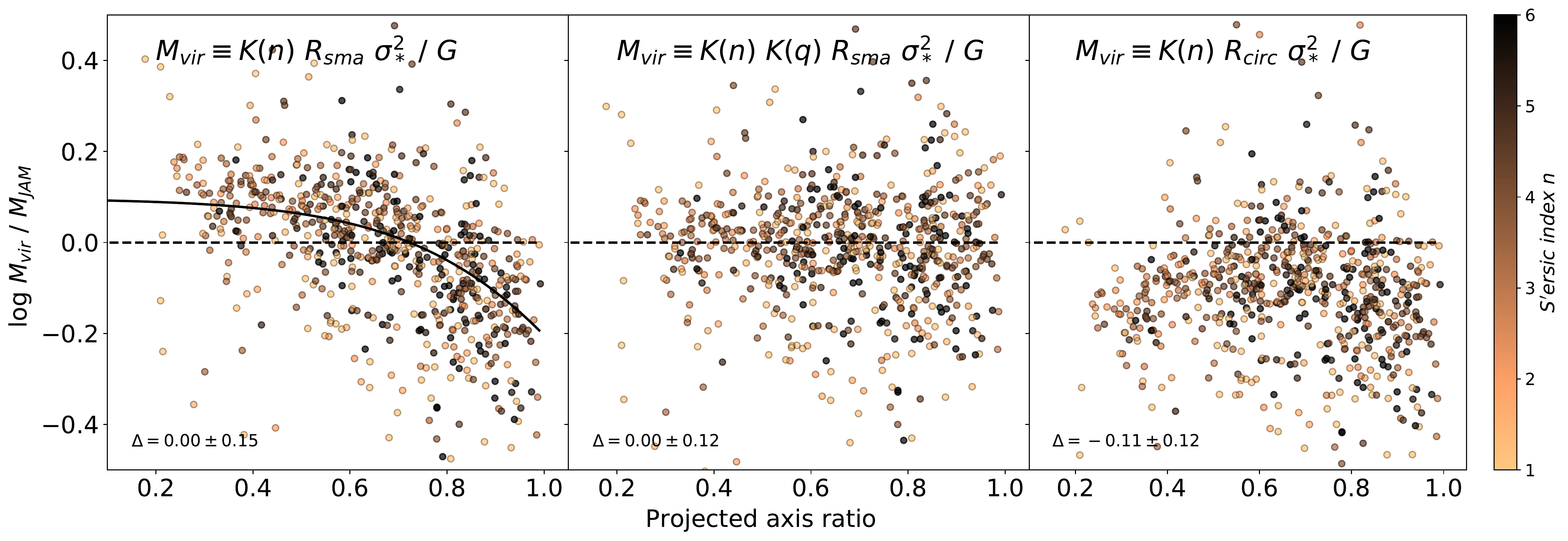} %rms_examples_paper
    \caption{The ratio between homology-corrected virial mass and Jeans mass for three different definitions of the virial mass (as indicated in the panels) versus projected axis ratio. \textit{Left:} $M_{\rm{vir}}$ with $R_{\rm{sma}}$ and $\sigpr$. The black line in the left-hand panel is a fit to the axis ratio trend and functions as the structural homology correction $K(q)$ (Eq.~\ref{eq:kq}). This correction is applied to $M_{\rm{vir}}$ in the \textit{middle} panel. 
    \textit{Right:} $M_{\rm{vir}}$ with $R_{\rm{circ}}$ instead of $R_{\rm{sma}}$ and without $K(q)$.\label{fig:sigma_correction}}
\end{figure*}

We define the total dynamical mass from the best-fitting Jeans models as 2$\times$ the Jeans model mass enclosed within a sphere with radius $R_{\rm{sma}}$, the semi-major axis of the ellipse that contains 50\% of the (projected) S\'ersic light model:
\begin{equation}
    M_{\rm{JAM}} \equiv 2\times M(r<R_{\rm{sma}}).
\end{equation}
This ensures that both mass estimates are approximately based on the same luminosity; essentially, our comparison is between mass-to-light ratios. 
The choice for $R_{\rm{sma}}$ is motivated in Section \ref{sec:radius}, which also includes a broader discussion on the concept of using any galactic radius as a proxy for virial radius.

As a starting point we calculate a simple virial mass estimate that is only proportional to $R_{\rm{sma}}$ and $\sigprsq$, the observed (projected) velocity second moment:

\begin{equation}
    M_{5} = 5\frac{\sigprsq R_{\rm{sma}}}{G}
\label{eq:mvir5}
\end{equation}
\noindent

The constant scaling factor 5 has been often used as a practical tool without explicit justification \citep[e.g.,][]{bender92, Jorgenson1996, VdWel2006, Toft2012}.  To place this normalization a firmer basis, C06 provided a calibration using detailed dynamical models as we do here. They found that, when the effective radius is measured in a then "classic" way using an $r^{1/4}$~growth-curve extrapolation \citep{dressler87, Jorgenson1996}, the best fitting coefficient was indeed $K=5.0\pm0.1$.

In Figure \ref{fig:mass_mass} we compare $M_5$ with $M_{\rm{JAM}}$.  There is a strong trend with S\'{e}rsic index $n$; clearly, galaxies are not self-similar in detail. In Figure \ref{fig:n_k} we explicitly show this $n$-dependence in comparison with the proportionality factor

\begin{equation}
K(n) = 8.87 - 0.831 n + 0.0241 n^{2}.
\label{eq:kn}
\end{equation}

\noindent
$K(n)$ is the scaling factor taken from C06 (Eq.~20).  The residual correlation between $M_5$ with $M_{\rm{JAM}}$ follows this description very well, indicating that this non-homology correction is required.
This agrees with the finding by \citet{Taylor2010}, who compared stellar masses from population with virial mass estimates. It also agrees with the results by C13, who used JAM dynamical models as we do here and also concluded that the above non-homology correction is needed when the effective radius is measured from S\'ersic models as we do here. However, we are now left with a strong residual correlation with projected axis ratio: round galaxies have larger $M_{\rm{JAM}} / M_{\rm{vir}}$ than flat galaxies,  signaling the importance of non-homology in 3D galaxy structure (in essence, spheres and disks) and the resulting projection effects on both the kinematics and the light distribution.

\subsection{Non-Homology in 3D Galaxy Structure and Projection Effects}
\label{sec:results_p1}
The projected axis ratio $q$ reflects the combination of the intrinsic, 3-dimensional geometry of a galaxy and the viewing angle. We use the residual trend with $q$ in Figure \ref{fig:n_k} to derive a second structural homology correction to account for variations in galaxy geometry that also includes projection effects. We do so purely empirically, by removing the residual trend with axis ratio $q$.

In the left-hand panel of Figure \ref{fig:sigma_correction} we show the same residual trend with $q$~as in Figure \ref{fig:n_k} but now explicitly as a function of $q$ and adopting the virial mass

\begin{equation}
    M_{\rm{n}} = K(n)\frac{\sigprsq R_{\rm{sma}}}{G}
\label{eq:mvirk}
\end{equation}
that includes the radial non-homology correction $K(n)$. With this definition, galaxies that are round in projection have underestimated virial masses; that is, the projected velocity second moment of nearly face-on, rotating galaxies do not `see' galactic rotation.  It is important to note here that the velocity dispersion used in the virial mass estimate must include all sources of motion in the galactic potential: not only the quasi-random motions associated with the true velocity dispersion at a given location in a galaxy, but also organized motions such as rotation. But a striking feature is that there is no systematic offset between $M_{\rm{n}}$ and $M_{\rm{JAM}}$. Apparently, variations in geometry and projection effects to do not cause a systematic difference between simple virial mass estimates and more accurate dynamical models using spatially resolved kinematics. 

We now introduce the homology correction $K(q)$:

\begin{equation}
K(q) = (0.87+0.38  e^{-3.78(1-q)})^2
\label{eq:kq}
\end{equation}
which is the inverse of the solid line in the left-hand panel of Figure \ref{fig:sigma_correction}.  This analytical form is purely practical and has no physical basis. For a given geometry and dynamical structure an inclination correction can be derived from the dynamical model or calculated directly from the tensor virial theorem \citep{bender92}, but our sample consists of a set of galaxies with a large variety in structure.

The middle panel of Figure \ref{fig:sigma_correction} shows the distribution of $M_{\rm{vir}}/M_{\rm{JAM}}$ according to our best-effort virial mass estimate:

\begin{equation}
    M_{\rm{n,q}} = K(n)K(q)\frac{\sigprsq R_{\rm{sma}}}{G}
\label{eq:mvir}
\end{equation}
\noindent
which is now independent of $q$ (by construction) and for which the variance is reduced by $\sim1/3$ (the new rms is 0.12 dex)\footnote{LEGA-C DR3 includes two quantities related to the stellar velocity dispersion: the measured stellar velocicity second moment {\tt SIGMA\_STARS\_PRIME} (written as $\sigpr$ in this paper) and {\tt SIGMA\_STARS\_VIR} = $\sqrt{K(q)} \times$~{\tt SIGMA\_STARS\_PRIME}.}. Importantly, no dependence on S\'ersic index is re-introduced: on average, the inclination correction works well for both high- and low-$n$ galaxies.

As mentioned before, the norm in the literature on virial mass estimates of high-redshift galaxies has been to choose the circularized radius $R_{\rm{circ}} \equiv \sqrt{q}R_{\rm{sma}}$. This is motivated by the result that the stellar-to-virial mass ratio produces smaller scatter when $R_{\rm{circ}}$ in a virial mass estimate,  both at low redshift (C13) and at high redshift \citep{Belli2017}, where the latter interpret this as evidence for rotational support via tentative residual correlations with axis ratio and S\'ersic index.

In the right-hand panel of Figure \ref{fig:sigma_correction} we show the result of using $R_{\rm{circ}}$ (and $K(n)$ but not $K(q)$). Compared to the left-hand panel there is a much weaker trend with axis ratio. The factor $\sqrt{q}$ -- replacing our $K(q)$ -- acts as a reasonably good homology correction, but this comes at the expense of a systematic offset. This offset is to be expected since the virial mass is now derived on the basis of a smaller radius than the Jeans model mass (by approximately a factor $\sqrt{q}$), but reducing the Jeans model mass by re-calculating it within a sphere of radius $R_{\rm{circ}}$ instead of $R_{\rm{sma}}$ would re-introduce the same axis ratio trend as seen in the left-hand panel.

\section{Discussion}
\label{section:Discussion}

\subsection{The Choice of Virial Radius}
\label{sec:radius}

Defining the virial radius of a galaxy is conceptually problematic. The only true virial radius is that of the dark matter halo, but this is not traced by the luminous body. In this paper, implicitly making several assumptions and approximations, we equate the virial radius with $R_{\rm{sma}}$, the semi-major axis of the ellipse that contains 50\% of the light in the HST ACS/F814W image, and we compare the inferred virial mass estimate with the Jeans model mass calculated within a {\it sphere} with the same radius. Specifically, we assume that: 1) the sphere with radius $R_{\rm{sma}}$ contains 50\% of the 3D luminosity distribution; 2) $R_{\rm{sma}}$, measured at a rest-frame wavelength of $\sim4000-5000$\AA~by fitting a 2D S\'ersic profile can be used as proxy for the spatial extent of galaxies.

The first of these assumptions was first addressed by \citet{ciotti91}, who showed that for a spherical galaxy with a S\'ersic profile $r_{1/2}/R_{\rm{sma}}=$1.34-1.36 for S\'ersic indices $n=2-10$ (here, $r_{1/2}$ is the radius of a sphere that contains 50\% of the 3D light distribution). But for disk galaxies this value decreases and approaches unity (C13). A key consideration is, then, that most galaxies ($\sim90\%$) in the LEGA-C sample analyzed in this paper are rotating, rather flat (intrinsic $c/a\approx 0.3$) and nearly axi-symmetric systems as evidenced by both their projected shape distribution \citep{Chang2013, VdWel2014b} and their kinematics (vH21). This large fraction of highly flattened galaxies is not specific to the LEGA-C sample: galaxies in the present-day Universe, including massive quiescent/early-type galaxies, generally have similar shapes \citep[e.g.,][]{Chang2013} and commonly show a large degree of rotational support \citep[e.g.,][]{Emsellem2011}. \citet{van-de-ven21} show that for such flattened, oblate galaxies the difference between the projected $R_{\rm{sma}}$ and $r_{1/2}$ of a sphere is negligible.  A minor caveat is that for slowly rotating triaxial galaxies (and more generally, galaxies with non-disklike geometries, in total about 10\% of the LEGA-C sample) $R_{\rm{sma}}$ and $r_{1/2}$ can differ: \citet{van-de-ven21} find that for massive, triaxial ellipticals $r_{1/2}/R_{\rm{sma}} = 1.18\pm0.18$, where the error reflects the galaxy-to-galaxy scatter due to variations in intrinsic shape and viewing angle.

These considerations generalizes the conclusions from previous work by \citet{Hopkins2010} and C13 who showed that $R_{\rm{sma}}$ is largely independent of inclination and is therefore the preferred size proxy (rather than $R_{\rm{circ}}$). We return to this issue in Section \ref{sec:atlas_sersic} when we examine the mass offset we see in Figure \ref{fig:sigma_correction} when using circularized radii (right-hand panel) and comparing with $M/L$ measurements from the ATLAS$^{\rm{3D}}$ survey of nearby early-type galaxies.

The second crucial assumption made explicit above (that the rest-frame $\sim4000-5000$\AA~$R_{\rm{sma}}$ from a S\'ersic fit is a good proxy for galaxy size) is more difficult to defend. The observed color gradients \citep[e.g.,][]{VdWel2012} imply that $R_{\rm{sma}}$ is wavelength dependent \citep[also see][]{kelvin12}, which in turn implies the presence of  mass-to-light gradients \citep{Szomoru2013, Mosleh2017, suess19}. If the inner parts of galaxies are dominated by stars, then the more sensible choice of the virial radius might be a mass-weighted half-light radius. At the same time, gas and dark matter fractions increase with radius, creating $M/L$ gradients in the opposite direction. Color gradient information is currently not available for the full sample of galaxies studied here. We should therefore keep in mind that our definition of the galaxy mass scale is set by our choice of $R_{\rm{sma}}$ as the optical half-light radius, measured at $\sim4000-5000$\AA, a choice that is to some extent arbitrary as it is determined by the available data.

In addition, our $R_{\rm{sma}}$ (and the stellar profile used in the Jeans dynamical model) relies on the S\'ersic profile. A comparison between S\'ersic model magnitudes and large-aperture ground-based photometric magnitudes convinces us that the S\'ersic profile is appropriate: the difference (accounting for differences in filter transmission curves) is, on average, just 0.02 mag, with 0.15 mag scatter. In particular, the S\'ersic model does not unduly extrapolate the light profile, artificially increasing the luminosity and the radius. We therefore believe the total luminosities to be accurate. This, in turn, implies that both $M_{\rm{JAM}}(R<R_{\rm{sma}})$ and $M_{\rm{vir}} / 2$ -- the approximate mass estimates within radius $R_{\rm{sma}}$ -- are accurate in relation to each other. The multiplication by a factor 2 is an unverified extrapolation and only serves to account for the total luminosity and to enable comparisons with, e.g., total stellar mass inferred from spatially integrated photometry.

Finally, for some purposes the circularized radius $R_{\rm{circ}}$ can be more useful. Traditionally, Fundamental Plane studies use $R_{\rm{circ}}$ and the projected axis ratio $q$ does not factor in. When $q$ is not available (for example, when sizes are derived from growth curves and circular apertures), then our $K(q)$ correction does not apply. Also, for extremely elongated, prolate galaxies $R_{\rm{circ}}$ is the more stable size proxy (compared to $r_{1/2}$). But overall, given the weak viewing angle dependence of $R_{\rm{sma}} / r_{1/2}$ for most galaxy geometries encountered in nature, we recommend the use of $R_{\rm{sma}}$ and the virial mass estimate from Eq.~\ref{eq:mvir}.

\begin{figure*}[t]
\epsscale{1.1}
\plotone{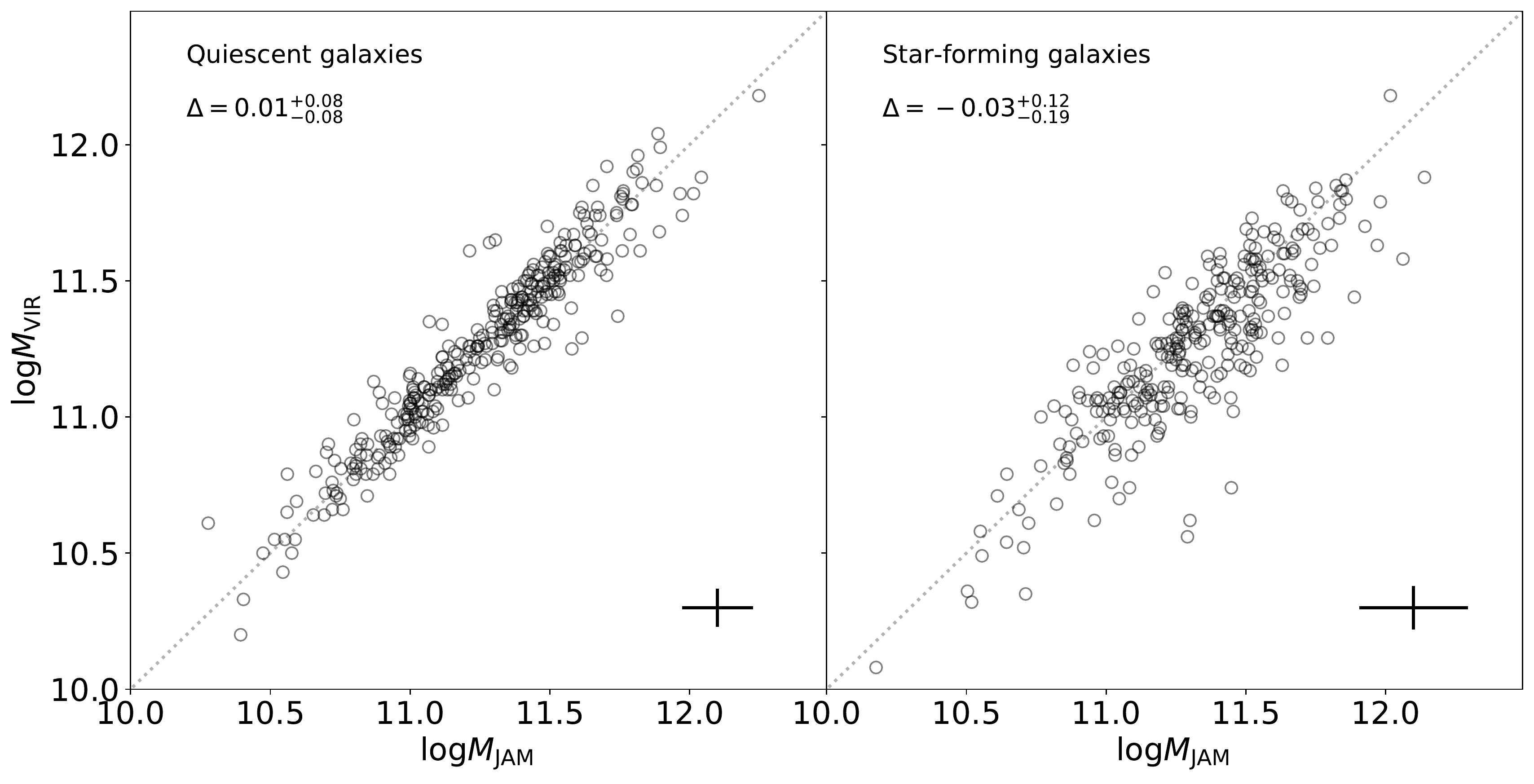} %rms_examples_paper
    \caption{Virial mass vs.~ JAM mass, separately for quiescent and star-forming galaxies. Our newly constructed virial mass agrees very well with the JAM mass, without systematic bias. The random uncertainty (as indicated by the scatter) is larger for star-forming galaxies than for quiescent galaxies.~\label{fig:jam_vir}}
\end{figure*}

\begin{figure}[t]
\epsscale{1.15}
\plotone{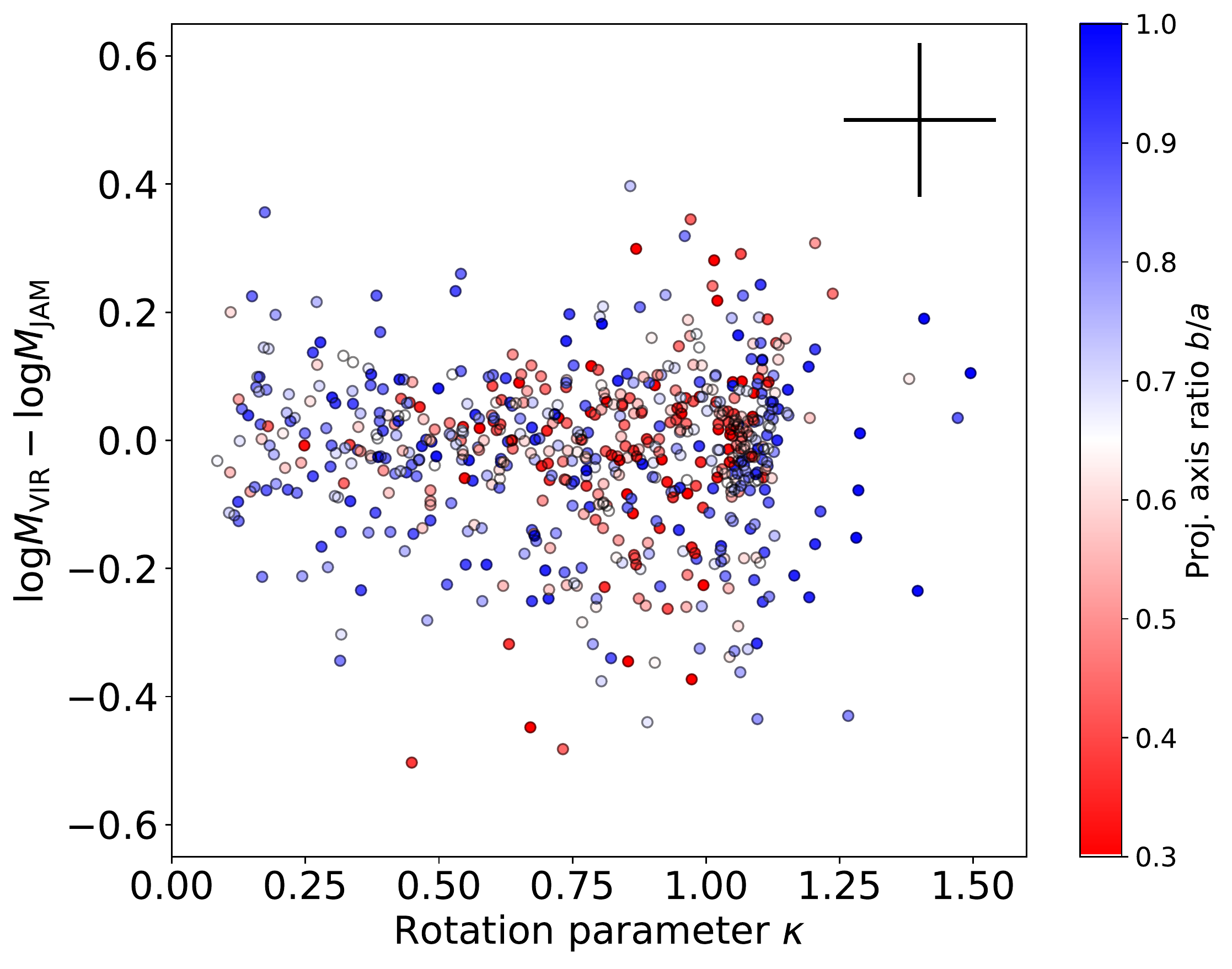} %rms_examples_paper
    \caption{$M_{\rm{vir}}/M_{\rm{JAM}}$ vs.~rotation parameter $\kappa$ calculated from the Jeans dynamical model, color-coded with projected axis ratio. $\kappa$ ranges from $\sim0$ for non-rotating galaxies to $\sim1$ for galaxies that are rotationally supported. Note that the dynamical model distinguishes between galaxies that are round in projection (color-coded with blue) that are face-on disks and and galaxies that are intrinsically round and non-rotating. There is no bias in our $M_{\rm{vir}}$ estimates that depends on the orbital structure of the galaxy; our inclination correction $K(q)$ serves to account for any such dependence.~\label{fig:rot_mass}}
\end{figure}

\subsection{Residual Correlations}
As discussed in Section \ref{section:results}, correlations in $M_{\rm{vir}}/M_{\rm{JAM}}$ with S\'{e}rsic $n$ and projected shape have been accounted for and removed in our final $M_{\rm{vir}}$ estimate. We find no significant correlations with any other parameter that is available for our sample. In particular, there is no difference between large and small galaxies, high- and low-mass galaxies, and no dependence on star-formation activity. Furthermore, we do not find a trend with redshift.

Most importantly, in Figure \ref{fig:rot_mass} we show that there is no residual correlation with the rotation parameter $\kappa$ derived from the Jeans models \citep[see][for details]{houdt21}. This is in contrast with the findings of \citet{WelMarel2008}, who find that fast-rotating galaxies overestimate the virial mass by as much as $0.2$ dex. However, that comparison is done at fixed $K=5$. As we show here, and demonstrated earlier for present-day galaxies (C13), a non-homologous scale factor should be used to derive unbiased masses, at least when using a S\'ersic profile-based effective radius. Using $K(n)=5$, we find a difference of at most $0.1$ dex between the fast rotating galaxies ($\kappa>0.5$) and slow rotating galaxies ($\kappa<0.5$). 
%This includes galaxies of all morphological types, however, whereas \citet{WelMarel2008} only study E, S0 and Sa type galaxies.
Whether or not this entirely explains the results from \citet{WelMarel2008} remains unclear, but this discrepancy is indicative of the importance of using consistent measurements when deciding which normalisation to use and when comparing galaxies across different epochs. \\ %discuss the influence on M/L evolution, like in Marel&Wel2008?

The newly calibrated virial mass estimates also apply equally, in a systematic sense to within 10\% or 0.04 dex, to quiescent and star-forming galaxies (Figure \ref{fig:jam_vir}). The sample is separated based on their rest-frame $U-V$~and $V-J$~colors. Both types show no systematic difference between $M_{\rm{vir}}$~and $M_{\rm{JAM}}$, but star-forming galaxies show larger scatter. 
%This is also consistent with the results from the MaNGA survey (\citet{LiMaoCappellari2018}). 
The scatter is consistent with the formal uncertainties in the $M_{\rm{JAM}}$~estimates. These are slightly larger than the formal uncertainties on $M_{\rm{vir}}$ due to the added flexibility provided by the dark-matter component in the JAM model. The uncertainties for the star-forming galaxies are larger than for the quiescent galaxies for several reasons: 1) the $\sigpr$~measurements are less precise as a result of lower $S/N$; 2) the stellar light profiles likely suffer from stronger deviations from the mass-follows-light assumption for the stellar component due to dust and star-forming regions; and 3) gas and dark matter fractions are likely higher.

\section{Comparison with ATLAS$^{\rm 3D}$}
\label{A3D}

The JAM-based mass models from the ATLAS$^{\rm 3D}$ survey (C13) have served as the standard benchmark for virial mass estimates. The motivation for our work is to directly determine the  normalization of the virial mass for galaxies at large lookback time, reducing potential observational biases and evolutionary effects. The self-consistent, mass-follows-light dynamical models for the ATLAS$^{\rm 3D}$ data from C13 take Multi-Gauss Expansion models as the light and mass tracer; S\'ersic profiles are fitted independently by \citet{Krajnovic2013} but are not used in the modeling. C13 provide two separate virial mass estimates estimates for the MGE light model (here, referred to as $M_{\rm{vir}}$) and the S\'ersic light model (here, referred to as $M_{\rm{vir,n}}$).  In Section \ref{sec:atlas_mge} we examine the residual trend with projected axis ratio in the $M_{\rm{vir}}$ estimates for the ATLAS$^{\rm 3D}$ sample, and in Section \ref{sec:atlas_sersic} we discuss a systematic difference of 0.1 dex  between LEGA-C and ATLAS$^{\rm 3D}$ mass estimates. 

\begin{figure*}
\epsscale{1}
\plotone{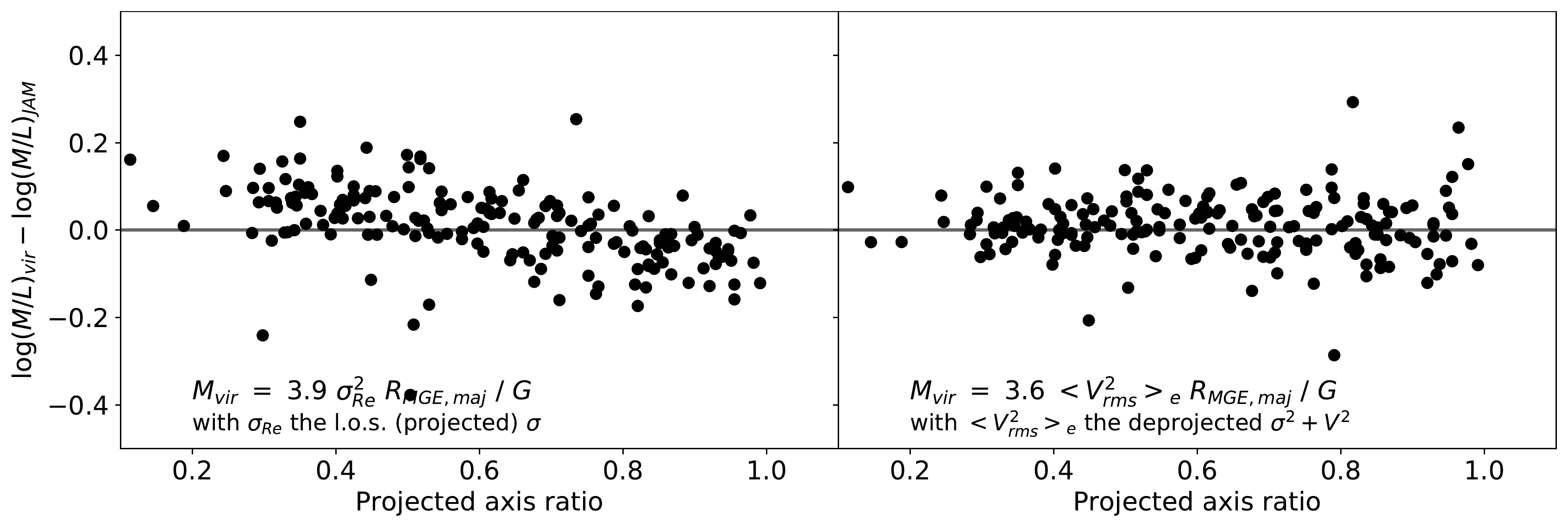}
\caption{Comparison of MGE-based virial $M/L$ with JAM $M/L$ for ATLAS$^{\rm 3D}$ from C13 as a function of projected axis ratio. Galaxies with quality flag 0 (see C13) are removed from the sample. \textit{Left:} $M_{\rm{vir}}/L$ as defined by C13 for MGE parameters; \textit{Right:} A revised $M_{\rm{vir}}/L$ for MGE parameters using the inclination correction velocity estimator $\langle V_{\rm rms}^2 \rangle_e$ from C13. See text for details. \label{fig:dml_q__mge_atlas3d}}
\end{figure*}

\begin{figure*}
\epsscale{1}
\plotone{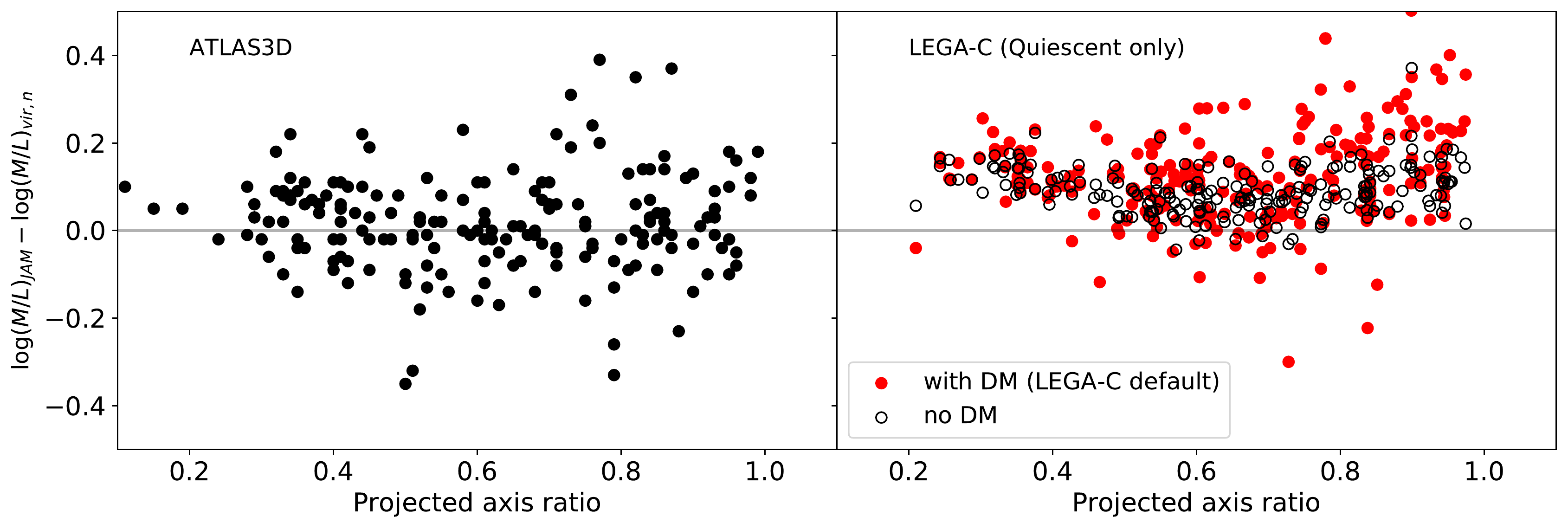}
\caption{Comparison of JAM $(M/L)$ from ATLAS$^{\rm 3D}$ (left) and LEGA-C (right) with the S\'ersic profile-based virial $M/L$ estimates as defined for ATLAS$^{\rm 3D}$ by C13 ($M_{\rm{vir,n}}=K(n)\sigma_{\rm{Re}}^2R_{\rm{Sersic,circ}}/G$).  For LEGA-C only quiescent galaxies are included to create a more direct comparison with the early-type ATLAS$^{\rm 3D}$ sample. For LEGA-C we show two flavors of JAM $M/L$. In red the $M/L$ from the default model with a dark matter component (calculated within $R_e$) and in black, for direct comparison with ATLAS$^{\rm 3D}$, the model without dark matter.
 \label{fig:legac_atlas3d}}
\end{figure*}

\subsection{A Dependence of the Virial Mass on Projected Axis Ratio in \rm{ATLAS$^{\rm 3D}$}}\label{sec:atlas_mge}

The left-hand panel of Figure \ref{fig:dml_q__mge_atlas3d} compares $(M/L)_{\rm{JAM}}$ and $(M/L)_{\rm{vir}}$ where both are based on the MGE light model, and where $M_{\rm{vir}}$ is defined in C13. This reveals a significant and hitherto hidden dependence on projected axis ratio, analogous to the trend seen for LEGA-C in this paper. With this new insight we define a new virial mass estimate for ATLAS$^{\rm 3D}$ (shown in the right-hand panel of Figure \ref{fig:dml_q__mge_atlas3d}): 

\begin{equation}
    M_{\rm{vir,A3D}} = 3.6\frac{\langle V_{\rm rms}^2 \rangle_e R_{\rm{MGE,maj}}}{G}
\label{eq:mvir_a3d}
\end{equation}
\noindent
where $\langle V_{\rm rms}^2 \rangle_e$ is the deprojected second moment of the velocity, from Eq. 29 of C13.\footnote{The values are available from https://purl.org/atlas3d}
This is the quadratic sum of the velocity dispersion $\sigma$ and the inclination-corrected velocity $V/\sin i$, averaging (light-weighted) over all spatial elements within the MGE half-light ellipse. Here, $i$ is the inclination inferred from the JAM. This inclination-corrected $\langle V_{\rm rms}^2 \rangle_e$ is not an empirical function of the observed axis ratio (as in Eq.~\ref{eq:kq} of this paper), but derives from the projected shape of the kinematics and the inclination, simultaneously removing the trend with axis ratio and reducing the scatter in the virial mass estimate. Specifically, the scatter in the ratio decreases from 0.08 dex in the left-hand panel to 0.06 dex in the right-hand panel. This result confirms one of the main findings of this paper for ATLAS$^{\rm 3D}$: the scatter in the virial estimates and the dependency on $q$ is partially due to the effect of inclination on the measured second velocity moment.  The normalization factor is reduced from 3.9 to 3.6 in order to remove an offset with respect to $(M/L)_{\rm{JAM}}$ (since generally $\langle V_{\rm rms}^2 \rangle_e > \sigma^2_{\rm{Re}}$). Note that no S\'ersic index dependence enters in the above: the mixed use of MGE light models for JAM and a S\'ersic index-based homology correction is generally not recommended (also see C13).

\subsection{A Systematic Offset between \rm{LEGA-C} and \rm{ATLAS$^{\rm 3D}$}}\label{sec:atlas_sersic}

C13 also provide a S\'ersic-based $M_{\rm{vir,n}}$ estimate, which uses $K(n)$ (here, Eq.\ref{eq:kn}) and $R_{\rm{circ}}$. We already saw in Section \ref{sec:results_p1} and Figure \ref{fig:sigma_correction} (right-hand panel) that this definition produces a mass offset with respect to the LEGA-C JAM estimates, suggesting a systematic difference between the LEGA-C and ATLAS$^{\rm 3D}$ mass scales, which we examine here.

As a first step we show for ATLAS$^{\rm 3D}$ that $(M/L)_{\rm{vir,n}}$ is independent of axis ratio (left-hand panel of Fig.~\ref{fig:legac_atlas3d}), reaffirming that using $R_{\rm{circ}}$ instead of $R_{\rm{sma}}$ serves as a first-order inclination correction on $M_{\rm{vir,n}}$. Note that the sign of the y-axis is reversed with respect to Figure \ref{fig:dml_q__mge_atlas3d} because we now wish to compare various JAM $M/L$ estimates to a common virial mass estimator ($M_{\rm{vir,n}}$ from C13). We also note that the scatter (0.12 dex) is larger than in Figure \ref{fig:dml_q__mge_atlas3d} because the S\'ersic models have different $R_e$ and $L$ compared to the MGE models that are used for both the JAM and the $M_{\rm{vir}}$ in Figure \ref{fig:dml_q__mge_atlas3d}.

For LEGA-C we have two JAM flavors: our preferred model, which we use in this paper, includes a dark matter component (necessitated by the relatively large extent of the kinematic data, typically $1-3R_e$), but vH21 also presented self-consistent mass-follows-light models, analogous to the ATLAS$^{\rm 3D}$ modeling results used in this section.
We calculate JAM $M/L$ for LEGA-C by integrating both the mass and the luminosity within a sphere with radius $R_{\rm{sma}}$. The LEGA-C mass-follows-light model $M/L$ shows a 0.09 dex offset w.r.t. $(M/L)_{\rm{vir,n}}$, whereas the model with dark matter produces a 0.12 dex offset. This implies that it is not the inclusion of a dark component that elevates the LEGA-C JAM $M/L$ by $\approx 0.1$ dex.

The other implication is that the offset in $M$ in the right-hand panel of Figure \ref{fig:sigma_correction} is due to an offset in $M/L$. This is not self-evident, as this proportionality rests on the assumption that the sphere with radius $R_{\rm{sma}}$ includes 50\% of the luminosity. This is not an exact equality, but for the LEGA-C JAM models the difference is small: the fraction of the luminosity within a sphere of radius $R_{\rm{sma}}$ is, on average, 0.47$\pm$0.02, where the error is the average random error on the individual estimates. This justifies our choice to compare the virial mass estimates with 2$\times$~the dynamical model mass calculated within a sphere with radius $R_{\rm{sma}}$ (see also Sec.~\ref{sec:radius}). 

The offset in $M/L$ implies that, for a given set of observables ($q$, $n$, $R_e$, and $\sigma$), $M_{\rm{JAM}}$ is 0.1 dex larger for LEGA-C than for ATLAS$^{\rm 3D}$.  It is not clear whether this offset is physical or the result of an unknown systematic error.  There are many differences between higher- and lower-redshift galaxies and between the ATLAS$^{\rm 3D}$ and LEGA-C samples. About 50\% of the LEGA-C sample are late-type galaxies, with presumably high gas and dark matter fractions relative to the early-type galaxies in ATLAS$^{\rm 3D}$. But also the early-type galaxies in LEGA-C are not equivalent to those in the ATLAS$^{\rm 3D}$ sample. The average $\sigma_{\star}^2$ is $>$2$\times$ higher for LEGA-C, and higher-redshift early-type galaxies are more compact and more rotation dominated. It is therefore not implausible that the dark matter fraction and overall structure is different.

On the pragmatic side, the LEGA-C stellar kinematic data for quiescent (early-type) galaxies typically probe out to 2$R_e$ for LEGA-C, which necessitates the inclusion of a dark matter component, whereas for ATLAS$^{\rm 3D}$ this is $1R_e$, for which models with and without dark matter produce very similar total $(M/L)(<R_e)$ (C13). It is possible that the LEGA-C $M/L(<R_e)$ estimates are biased upward by the statistical weight of kinematic data outside $R_e$ in combination with the low spatial resolution. Kinematic data with higher spatial resolution from, e.g., ELT is required to resolve this issue. Remaining agnostic about the interpretation of the offset in mass scale between LEGA-C and ATLAS$^{\rm 3D}$, we conclude that we have established a common mass scale for galaxies across 8 Gyr of cosmic time with a small systematic uncertainty of 0.1 dex.

\section{Summary \& Conclusions}
\label{section:conclusions}

In this paper we provide a new calibration of the mass scale of galaxies at $z=0.6-1$ that is applicable to galaxies of all morphological types.  Jeans axi-symmetric models for 673 galaxies based on spatially resolved long-slit stellar kinematics from LEGA-C serve as the baseline. Integrated stellar velocity dispersions (the second moment) and S\'ersic profiles from HST imaging then allow for virial mass estimates (Eq.~\ref{eq:mvir}) with a systematic uncertainty with respect to the locally calibrated mass scale of at most 0.1 dex and with 20\% random uncertainty for quiescent galaxies and 40\% random uncertainty for star-forming galaxies (Figure \ref{fig:jam_vir}). The combination of elements to arrive at this level of consistency are as follows:

\begin{itemize}
    \item {\bf Non-homology in radial structure} as parameterized in Eq.~\ref{eq:kn} in order to remove any dependence on galaxy structure (S\'ersic index $n$). Without this correction, and adopting a standard proportionality factor of $K=5$, disk galaxies will have their dynamical masses underestimated by $>50\%$~(Figure \ref{fig:mass_mass}). We note that the use of $K(n)$ is contingent on the use of the S\'ersic profile as a proxy for the light profile, as demonstrated previously by \citet{Cappellari2013+jam}.
    \item {\bf Non-homology in 3D galaxy structure} as parameterized in Eq.~\ref{eq:kq} in order to remove any dependence on projected axis ratio. This accounts for the combined effect of variations in intrinsic, 3D galaxy shape and projection effects. Without such a correction, using the measured, projected velocity second moment (referred to as $\sigpr$ in this paper), face-on galaxies will have underestimated masses, and edge-on galaxies overestimated masses (Figure \ref{fig:sigma_correction}, left-hand panel).
    \item {\bf Half-light radius} as measured along the major axis ($R_{\rm{sma}}$), as previously demonstrated by \citet{Cappellari2013+jam}.
\end{itemize}

For convenience we repeat the relevant equations here. The calibrated virial mass estimate is defined as:
\begin{equation}
    M_{\rm{vir}} = K(n)K(q)\frac{\sigprsq R_{\rm{sma}}}{G}
\end{equation}
\noindent where
\begin{equation}
K(n) = 8.87 - 0.831 n + 0.0241 n^{2} \rm{(from C13)}
\end{equation}
\noindent and
\begin{equation}
K(q) = (0.87+0.38  e^{-3.78(1-q)})^2
\end{equation}
\noindent with the projected axis ratio $q\equiv b/a$. 

A comparison with the low-redshift sample of early-type galaxies with IFU data from ATLAS$^{\rm{3D}}$ shows that the LEGA-C dynamical masses are systematically higher by 0.1 dex (Section \ref{sec:atlas_sersic}). It is not clear whether this offset is due to structural differences between the galaxies in the two samples, or due to an unknown systematic error. Nonetheless, we stress that a common mass scale for galaxies across 8 Gyr of cosmic time with a systematic uncertainty of at most 0.1 dex should be considered a success.

Numerous applications, spin-offs and expansions are possible. Practical applications of our calibrated virial mass scale include quick, unbiased dynamical mass estimates for large samples with rudimentary measures of the velocity second moment and size, and the cross-comparison of dynamical masses based on measurements from different instruments\footnote{Subtle differences in integrated $\sigma$ measurements can be expected when comparing slit- and fiber-based spectra, but at higher redshift the apertures are sufficiently large to render potential aperture corrections small compared to the uncertainties on the mass estimates. } Future work will include a one-to-one comparison between dynamical models based on ionized gas kinematics and stellar kinematics at large look-back time; a comparison with stellar mass estimates, augmented with either direct or inferred gas mass estimates; separating the radial dependence of stellar $M/L$, gas mass fractions and dark matter fraction (for a smaller subset with highly significant deviations from the mass-follows-light assumption), and a comparison with total masses of galaxies in cosmological hydrodynamical simulations.

\section*{Acknowledgments}
Based on observations made with ESO Telescopes at the La Silla Paranal Observatory under programme ID 194-A.2005 (The LEGA-C Public Spectroscopy Survey). This project has received funding from the European Research Council (ERC) under the European Union’s Horizon 2020 research and innovation programme (grant agreement No. 683184).

\bibliographystyle{aasjournal}
\bibliography{ms_v2}

\restartappendixnumbering

\appendix{}
\label{AppendixA}

In this Appendix we provide, as a supplement, the dynamical model parameters presented by vH21. The virial mass values were published by \citet{vanderwel21}. In vH21 we already published mass estimates and proxies for dynamical structure (e.g., $V/\sigma$), but to facilitate comparisons with other datasets and models we present here all fitted parameters, including those without precise constraints (essentially, nuisance parameters required only to marginalize over astrophysically motivated priors). For a full description of the models we refer to vH21, and here we only repeat the important caveats that should be kept in mind when using these data. 

In Table \ref{tab:cat1} we include the $M/L$ estimates for the mass-follows-light model as well as the model that includes a dark component: in addition to the stellar component (as traced by the HST ACS/F814W images) a dark matter component is incorporated. This component serves to account for any $M/L$ gradient, regardless of whether it is actual dark matter, or rather due to gas or stellar $M/L$ variations.  While poorly constrained for most individual galaxies (this is reflected in the uncertainties in $f_{\rm{DM}}$, which are much larger than the uncertainties in the total $M/L$) we systematically find for the ensemble that a marginally positive $f_{\rm{DM}}$ is required to model our kinematic data that probe out to $1-3R_e$. In short, constraints on $f_{\rm{DM}}$ reflect a deviation from the mass-follows-light assumption and should not be taken at face value: careful analysis of individual galaxies is required to interpret these quantities. 

Inclinations are almost always unconstrained by the kinematic data due to the relatively poor spatial resolution and large width of the slit. The estimates in the table are therefore the result of the marginalization over the prior, which is set by the projected shape of the HST light profile (see vH21 for further details).  

%For completeness, we also present in Table \ref{tab:cat2} the results from the mass-follows-light models (producing a constant $M/L$), without a dark matter component. These $M/L$ estimates have underestimated uncertainties, and biases as a function of galaxy type and radial extent of the kinematic data are to be expected.

%Finally, the S\'ersic models used in this paper are published in \citet{vanderwel21} and the full MGE models can be downloaded on this website (http...)

\begin{deluxetable*}{cccccccc}
\tabletypesize{\scriptsize}
\tablecolumns{12}
%\tablewidth{0pt}
\tablecaption{LEGA-C JAM parameters (models with dark component) \label{tab:cat1}}
\tablehead{
\colhead{ID1} &  \colhead{ID2} &  \colhead{$\log{(L)}$} & \colhead{$\log{(M/L)}$~(no DM)} &  \colhead{$\log{(M/L)}$~(with DM)} & \colhead{$f_{\rm{DM}}$} & \colhead{$\beta_{\rm{z}}$} & \colhead{$i$}\\
\colhead{} & \colhead{} & \colhead{$L_{\odot\rm{,g}}$} & \colhead{$M_{\odot}/L_{\odot\rm{,g}}$} &  \colhead{$M_{\odot}/L_{\odot\rm{,g}}$}  & \colhead{} & \colhead{}  & \colhead{deg.} 
}
\startdata
5 & 4792 & 10.87 & $0.05^{+0.05}_{-0.03}$ & $0.34^{+0.27}_{-0.24}$ & $0.62^{+0.38}_{-0.42}$ & $0.03^{+0.33}_{-0.35}$ & $28\pm4$ \\
26 & 10462 & 10.73 & $-0.24^{+0.06}_{-0.05}$ & $0.15^{+0.46}_{-0.36}$ & $0.71^{+0.29}_{-0.58}$ & $0.02^{+0.32}_{-0.36}$ & $44\pm5$ \\
27 & 10902 & 11.35 & $0.09^{+0.05}_{-0.03}$ & $0.43^{+0.48}_{-0.42}$ & $0.62^{+0.38}_{-0.52}$ & $-0.03^{+0.34}_{-0.33}$ & $31\pm4$ \\
38 & 14375 & 11.25 & $0.14^{+0.02}_{-0.02}$ & $0.17^{+0.07}_{-0.08}$ & $0.10^{+0.76}_{-0.09}$ & $-0.07^{+0.34}_{-0.29}$ & $74\pm5$ \\
39 & 14729 & 11.37 & $0.47^{+0.05}_{-0.04}$ & $0.42^{+0.32}_{-0.40}$ & $0.00^{+1.00}_{-0.00}$ & $0.19^{+0.23}_{-0.41}$ & $67\pm6$ \\
\vdots & \vdots & \vdots & \vdots & \vdots & \vdots & \vdots
\enddata
%\vspace{-0.8cm}
\tablecomments{JAM model parameters described by vH21. (1): LEGA-C ID (DR3); (2):Ultra-VISTA ID \citep{Muzzin2013b}; (3): $g$ band Mass-to-light from the mass follows light model; (4): Total $g$ band mass-to-light ratio from the light plus dark matter model; (5): Dark matter fraction; 
(6): Inclination; (7): Vertical anistropy $\beta_{z}\equiv 1 - \langle v_{z}^{2}\rangle/\langle v_{R}^{2}\rangle$. All quantities are calculated within spherical apertures with radius $R_{\rm{Sersic,sma}}$. Values and uncertainties are based on 16th, 50th, 84th percentiles of posterior parameter distributions. The machine-readable table has 861 entries and is matched to the catalog published in vH21.} 
\end{deluxetable*}

\end{document}